\documentclass{ws-procs9x6-cpt22}
\begin{document}

\newcommand{\refeq}[1]{(\ref{#1})}
\def\etal {{\it et al.}}
\newcommand{\nup}{$\nu_{\pi}$}
%any other macros go here
%Particle symbols
\newcommand{\eb}{\mbox{$\overline{\mathrm e}$}}
\newcommand{\el}{\mbox{e$^{-}$}}
\newcommand{\hb}{\mbox{$\overline{\mathrm H}$}}
\newcommand{\mum}{\mbox{$\mu^{-}$}}
\newcommand{\muplus}{\mbox{$\mu^{+}$}}
\newcommand{\pos}{\mbox{e$^{+}$}}
\newcommand{\pbar}{\mbox{$\overline{\mathrm p}$}}
\newcommand{\Hbar}{\mbox{$\overline{\mathrm H}$}}
\newcommand{\meta}{\mbox{$\overline{\mathrm{p}}\mathrm{He}^{+}$}}
\newcommand{\Mpbar}{\mbox{$M_{\overline{\mathrm p}$}}}
\newcommand{\Qpbar}{\mbox{$Q_{\overline{\mathrm p}$}}}

%pbarhelium
\newcommand{\pbhe}{$\overline{\mathrm p} {\rm He}^{+}$}
\newcommand{\pbheion}{$\overline{\mathrm p}{\rm He}^{++}$}
%Vectors
\newcommand{\bv}{\mbox{\boldmath$B$}}
\newcommand{\ev}{\mbox{\boldmath$E$}}
%Special symbols
\newcommand{\degr}{\mbox{$^{\circ}$}}
\newcommand{\dul}[1]{\underline{\underline #1}}
\newcommand{\half}{\mbox{$\frac{1}{2}$}}
\newcommand{\mrm}{\mathrm}
\newcommand{\ol}{\overline}
\newcommand{\ul}{\underline}
\newcommand{\sss}{\mbox{$S_{1}$}}
%Arrows
\newcommand{\la}{\mbox{$\leftarrow$}}
\newcommand{\La}{\mbox{$\Leftarrow$}}
\newcommand{\Lra}{\mbox{$\Leftrightarrow$}}
\newcommand{\lra}{\mbox{$\leftrightarrow$}}
\newcommand{\Ra}{\mbox{$\Rightarrow$}}
\newcommand{\ra}{\mbox{$\rightarrow$}}
\newcommand{\da}{\mbox{$\downarrow$}}
\newcommand{\Da}{\mbox{$\Downarrow$}}
\newcommand{\ua}{\mbox{$\uparrow$}}
%Molecules
\newcommand{\hh}{\mbox{H$_{2}$}}
\newcommand{\nn}{\mbox{N$_{2}$}}
% Eberhard's
\newcommand{\pbhef}{$\overline{\mathrm p} ^4{\rm He}^{+}$}
\newcommand{\pbhet}{$\overline{\mathrm p} ^3{\rm He}^{+}$}
%% vectors schould be bold, spin and angular momentum lower case
%\newcommand{\Se}{$\vec{S}_e$}
%\newcommand{\Lp}{$\vec{L}_{\overline{p}}$}
%\newcommand{\Sp}{$\vec{S}_{\overline{p}}$}
%\newcommand{\Sh}{$\vec{S}_{h}$}
\newcommand{\Se}{$\boldsymbol{s}_\mathrm{e}$}
\newcommand{\Lp}{$\boldsymbol{l}_\mathrm{p}$}
\newcommand{\Sp}{$\boldsymbol{s}_\mathrm{p}$}
\newcommand{\Sh}{$\boldsymbol{s}_\mathrm{h}$}
\newcommand{\Lps}{$L_{\overline{p}}$}
%\newcommand{\mup}{$\mbox{\boldmath$\mu$}_{\overline{p}}$}
%% masses
%% trouble with hepnames for FFK2021
\newcommand{\mpbar}{$m_{\overline{\mathrm{p}}}$}
\newcommand{\mprot}{$m_{\mathrm{p}}$}
\newcommand{\mel}{$m_{\mathrm{e}}$}
\newcommand{\mpos}{$m_{\mathrm{e^+}}$}
%% magentic moments
\newcommand{\mupbar}{$\mbox{$\mu$}_{\APproton}$}
\newcommand{\msigma}{m$_{\PSigmaminus}$}
\newcommand{\musigma}{$\mbox{$\mu$}_{\PSigmaminus}$}
\newcommand{\mupscal}{$\mu_{\overline{p}}$}
\newcommand{\mups}{$\mbox{\boldmath$\mu$}_{\overline{p}}^{\mathrm{s}}$}
\newcommand{\mupsscal}{$\mu_{\overline{p}}^{\mathrm{s}}$}
\newcommand{\mupl}{$\mbox{\boldmath$\mu$}_{\overline{p}}^{\ell}$}
\newcommand{\mue}{$\mbox{$\mu$}_\mathrm{e}$}
\newcommand{\mup}{$\mbox{$\mu$}_\mathrm{p}$}
\newcommand{\muh}{$\mbox{\boldmath$\mu$}_{h}$}
\newcommand{\muhscal}{$\mu_{h}$}
\newcommand{\gh}{$g_{h}$}
\newcommand{\muB}{$\mu_{\mathrm B}$}
\newcommand{\muN}{$\mu_{\mathrm N}$}
\newcommand{\muNbar}{$\mu_{\overline{\mathrm N}}$}
\newcommand{\Dthexp}{$\Delta_{\mathrm{th-exp}}$}
\newcommand{\dexp}{$\delta_{\mathrm{exp}}$}
\newcommand{\Fp}{$F^+$}
\newcommand{\Fm}{$F^-$}
\newcommand{\fp}{$f_+$}
\newcommand{\fm}{$f_-$}
\newcommand{\tone}{$t_1$}
\newcommand{\ttwo}{$t_2$}
\newcommand{\gl}{$g^{\overline{p}}_{\ell}$ }
\newcommand{\gs}{$g^{\overline{p}}_{\mathrm{s}}$ }
\newcommand{\nuHFbar}{$\nu_{\mathrm{HF}}({\overline{\mathrm{H}}})$}
\newcommand{\nuHF}{$\nu_{\mathrm{HF}}$}
\newcommand{\EHF}{$E_{\mathrm{HF}}$}
\newcommand{\DnuHF}{$\Delta \nu_{\mathrm{HF}}$}
\newcommand{\nuHFp}{$\nu_{\mathrm{HF}}^+$}
\newcommand{\nuHFm}{$\nu_{\mathrm{HF}}^-$}
\newcommand{\nuHFpm}{$\nu_{\mathrm{HF}}^\pm$}
\newcommand{\nuSHF}{$\nu_{\mathrm{SHF}}$}
\newcommand{\nuSHFp}{$\nu_{\mathrm{SHF}}^+$}
\newcommand{\nuSHFpm}{$\nu_{\mathrm{SHF}}^\pm$}
\newcommand{\nuSHFm}{$\nu_{\mathrm{SHF}}^-$}
\newcommand{\nuMW}{$\nu_{\mathrm{MW}}$}

\newcommand{\nus}{$\nu_{\sigma}$}
\newcommand{\nupone}{$\nu_{\pi_1}$}
\newcommand{\nuptwo}{$\nu_{\pi_2}$}

\newcommand{\rprot}{$r_{\mathrm{p}}$}
\newcommand{\rZ}{$r_{\mathrm{Z}}$}
\newcommand{\dpol}{$\delta_{\mathrm{pol}}$}
\newcommand{\nuS}{$\nu_{\mathrm{1S-2S}}$}
\newcommand{\nuSH}{$\nu_{\mathrm{1S-2S}}^{\mathrm{H}}$}
\newcommand{\nuSHbar}{$\nu_{\mathrm{1S-2S}}^{\overline{\mathrm{H}}}$}
\newcommand{\nuLS}{$\nu_{\mathrm{2S-2P}}$}
\newcommand{\nuLSH}{$\nu_{\mathrm{2S-2P}}^{\overline{H}}$}
\newcommand{\nuLSHbar}{$\nu_{\mathrm{2S-2P}}^{\overline{\mathrm{H}}}$}
\newcommand{\nuHFS}{$\nu_{\mathrm{HFS}}$}
\newcommand{\nuHFSH}{$\nu_{\mathrm{HFS}}^{\overline{H}}$}
\newcommand{\nuHFSHbar}{$\nu_{\mathrm{HFS}}^{\overline{\mathrm{H}}}$}

\newcommand{\rpe}{$r_{\mathrm{p}}^{\mathrm{e}}$}
\newcommand{\ket}[1]{|#1\rangle}

\newcommand{\s}{\mbox{$\sigma_1$}}
\newcommand{\p}{\mbox{$\pi_1$}}
\newcommand{\ptwo}{\mbox{$\pi_2$}}
\newcommand{\Bext}{\mbox{$B_{\mathrm{ext}}$}}
\newcommand{\Bosc}{\mbox{$B_{\mathrm{osc}}$}}
\newcommand{\kh}{\mbox{$\hat{\kappa}$}}
\newcommand{\be}{\mbox{$\tilde{b}_3^e$}}
\newcommand{\bp}{\mbox{$\tilde{b}_3^p$}}
\newcommand{\nuab}{\mbox{$\nu_{\mathrm{ab}}$}}
\newcommand{\nuac}{\mbox{$\nu_{\mathrm{ac}}$}}
\newcommand{\nucd}{\mbox{$\nu_{\mathrm{cd}}$}}
\newcommand{\nuad}{\mbox{$\nu_{\mathrm{ad}}$}}

\def\nr{\rm NR}
\def\nrtemplate#1#2#3{#1^{\nr#3}_{#2}}

\def\anrf#1#2{\nrtemplate{{a_{#1}}}{#2}{}}
\def\cnrf#1#2{\nrtemplate{{c_{#1}}}{#2}{}}
\def\gzBnrf#1#2{\nrtemplate{{g_{#1}}}{#2}{(0B)}}
\def\goBnrf#1#2{\nrtemplate{{g_{#1}}}{#2}{(1B)}}
\def\goEnrf#1#2{\nrtemplate{{g_{#1}}}{#2}{(1E)}}
\def\HzBnrf#1#2{\nrtemplate{{H_{#1}}}{#2}{(0B)}}
\def\HoBnrf#1#2{\nrtemplate{{H_{#1}}}{#2}{(1B)}}
\def\HoEnrf#1#2{\nrtemplate{{H_{#1}}}{#2}{(1E)}}
%% EW
\def\Knr#1#2{\nrtemplate{{\mathcal{K}_{#1}}}{#2}{}}
\def\Knrlab#1#2{\nrtemplate{{\mathcal{K}_{#1}}}{#2}{,lab}}
\def\Knrsun#1#2{\nrtemplate{{\mathcal{K}_{#1}}}{#2}{,sun}}
\def\Vnr#1#2{\nrtemplate{{\mathcal{V}_{#1}}}{#2}{}}
\def\cnr#1#2{\nrtemplate{{c_{#1}}}{#2}{}}
\def\anr#1#2{\nrtemplate{{a_{#1}}}{#2}{}}
\def\TqPnr#1#2{\nrtemplate{{\mathcal{T}_{#1}}}{#2}{(qP)}}
\def\ToBnr#1#2{\nrtemplate{{\mathcal{T}_{#1}}}{#2}{(0B)}}
\def\TeBnr#1#2{\nrtemplate{{\mathcal{T}_{#1}}}{#2}{(1B)}}
\def\TEBnr#1#2{\nrtemplate{{\mathcal{T}_{#1}}}{#2}{(1E)}}
\def\gqPnr#1#2{\nrtemplate{{g_{#1}}}{#2}{(qP)}}
\def\HqPnr#1#2{\nrtemplate{{H_{#1}}}{#2}{(qP)}}

% CPT
\newcommand{\Csym}{$\mathcal{C}$}
\newcommand{\Psym}{$\mathcal{P}$}
\newcommand{\Tsym}{$\mathcal{T}$}
\newcommand{\CP}{$\mathcal{CP}$}
\newcommand{\CPT}{$\mathcal{CPT}$}
% SME
\newcommand{\Et}{$\tilde{E_0}$}
\newcommand{\Ete}{$\tilde{E_0}^\mathrm{e}$}
\newcommand{\Epe}{$\tilde{E_0}^\mathrm{p}$}
\newcommand{\bt}{$\tilde{b_0}$}
\newcommand{\bte}{$\tilde{b_0}^\mathrm{e}$}
\newcommand{\btp}{$\tilde{b_0}^\mathrm{p}$}
\newcommand{\btj}{$\tilde{b_j}$}
\newcommand{\btje}{$\tilde{b_j}^\mathrm{e}$}
\newcommand{\btjp}{$\tilde{b_j}^\mathrm{p}$}
% non-minimal SME
\def\nr{\rm NR}
\def\nrtemplate#1#2#3{#1^{\nr#3}_{#2}}

\def\anrf#1#2{\nrtemplate{{a_{#1}}}{#2}{}}
\def\cnrf#1#2{\nrtemplate{{c_{#1}}}{#2}{}}
\def\gzBnrf#1#2{\nrtemplate{{g_{#1}}}{#2}{(0B)}}
\def\goBnrf#1#2{\nrtemplate{{g_{#1}}}{#2}{(1B)}}
\def\goEnrf#1#2{\nrtemplate{{g_{#1}}}{#2}{(1E)}}
\def\HzBnrf#1#2{\nrtemplate{{H_{#1}}}{#2}{(0B)}}
\def\HoBnrf#1#2{\nrtemplate{{H_{#1}}}{#2}{(1B)}}
\def\HoEnrf#1#2{\nrtemplate{{H_{#1}}}{#2}{(1E)}} 

\title{In-beam hyperfine spectroscopy of antihydrogen, hydrogen and deuterium}

\author{E.\ Widmann$^1$ }

\address{$^1$Stefan Meyer Institute, Austrian Academy of Sciences, 1030 Vienna, Austria}

%\address{$^2$Group, Laboratory,\\
%City, State ZIP/Zone, Country}

\author{On behalf of the ASACUSA Collaboration\footnote{\url{https://cern.ch/asacusa}}}

\begin{abstract}
The ASACUSA  collaboration is developing a polarized beam of antihydrogen atoms to precisely determine the ground-state hyperfine structure  for studies of CPT and Lorentz invariance. Using a beam of ordinary hydrogen, measurements of both the $\sigma$ and $\pi$-transition have been performed, investigating orientation-dependent SME coefficients. Furthermore a first hyperfine experiment with a beam of deuterium is being prepared.
\end{abstract}

\bodymatter

\section{Introduction}
The ground-state hyperfine structure (GS-HFS) of antihydrogen offers a very sensitive test of CPT invariance by comparing it to the value for hydrogen which is known to a relative precision of $10^{-12}$. The GS-HFS is caused by the spin-spin interaction of electron/positron and proton/antiproton. In terms of absolute accuracy a measurement of the transition frequency \nuHFbar\ in antihydrogen with similar precision than in H could provide the most sensitive test of CPT in \Hbar\ \cite{Widmann:22}. For H, the finite magnetization distribution of the proton leads to corrections to the leading order term at a level of $\sim 30$ ppm, therefore a measurement of \nuHFbar\ to 1 ppm precision, which constitutes the first goal of the ASACUSA collaboration, will reveal the magnetic substructure of the antiproton \cite{Widmann:22}.

Within the SME framework, of the two primary transitions accessible in a Rabi-type experiment, the $\sigma$-transition ($(F,M_F)=(1,0)\rightarrow(0,0)$, $F$: total angular momentum, $M_F$: its projection) with frequency \nus\ is not sensitive to CPT nor Lorentz invariance violation, only the  $\pi$-transition ($(F,M_F)=(1,1)\rightarrow(0,0)$, frequency \nup). Measurements of siderial variations of \nup\ using a H maser \cite{Phillips:01,Humphrey:2003} have already put mHz precision limits on minimal SME coefficients, but within the non-minimal SME additional orientation-dependent coefficients appear that can be constrained in H experiments.

No measurements of SME coefficients have been performed so far in deuterium. Here, in the non-minimal SME large sensitivity gains are expected due to the dependence of corrections on powers of the relative momentum of proton and neutron in the d nucleus. 

\section{Status of the antihydrogen experiment}

The goal of ASACUSA is to generate a beam of polarized \Hbar\ by mixing \pbar\ and \pos\ in a so-called Cusp trap \cite{Mohri:2003wu} for a GS-HFS measurement. Several milestones towards this goal have been reached, but the achieved rate and ground-state fraction of the produced \Hbar\ \cite{KolbingerEtAl2015} is not yet sufficient for spectroscopy. However, recently the plasma temperature has been significantly reduced \cite{Hunter:22}, which is expected to both drastically increase the formation rate and ground-state fraction \cite{Radics:14}. These methods are being applied to the experiments currently under way at the AD/ELENA facility at CERN.

\section{Measurements using a hydrogen beam}

\begin{figure}
\begin{center}
\includegraphics[width=1.6in]{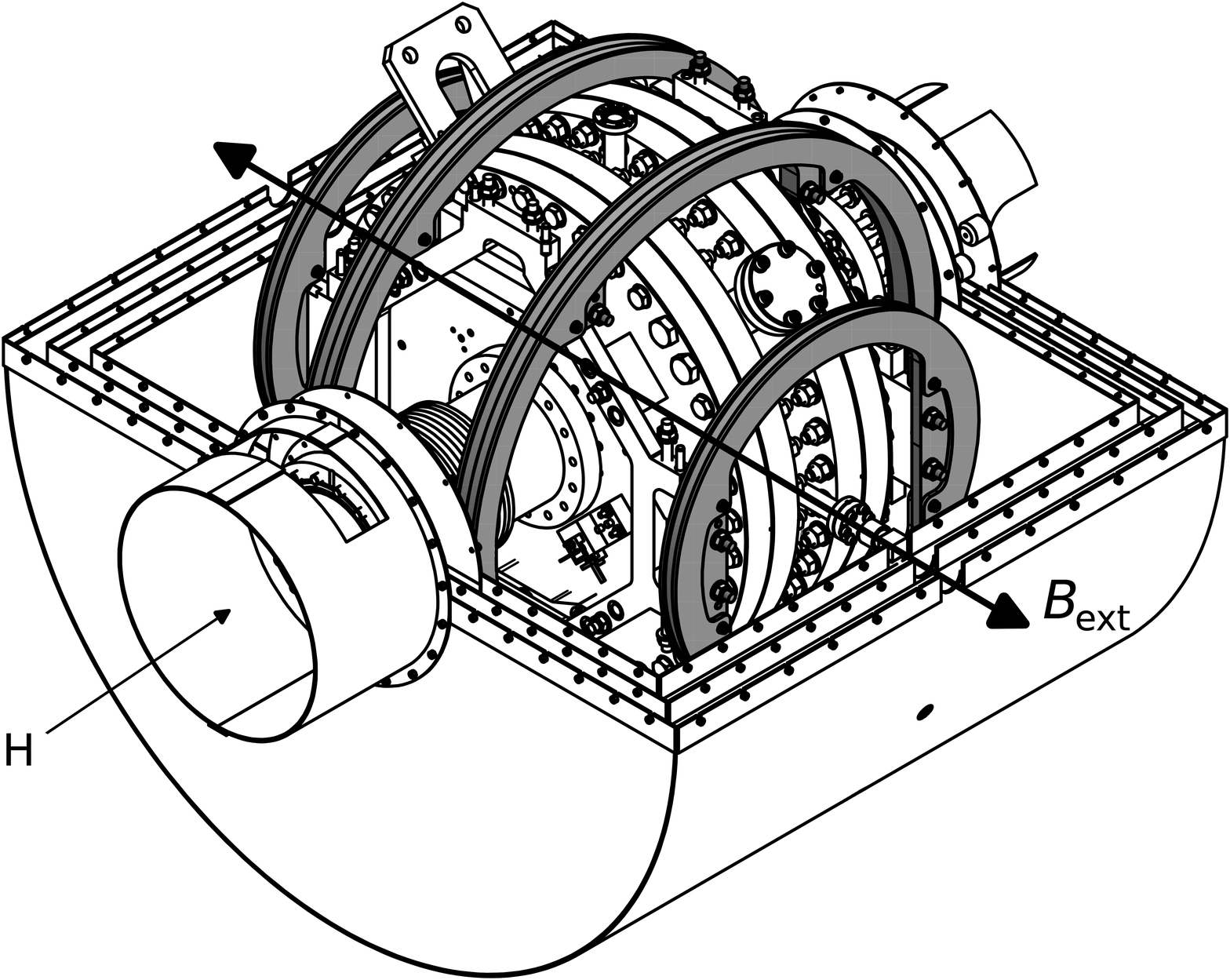} \hspace*{3mm}
\includegraphics[width=2.2in]{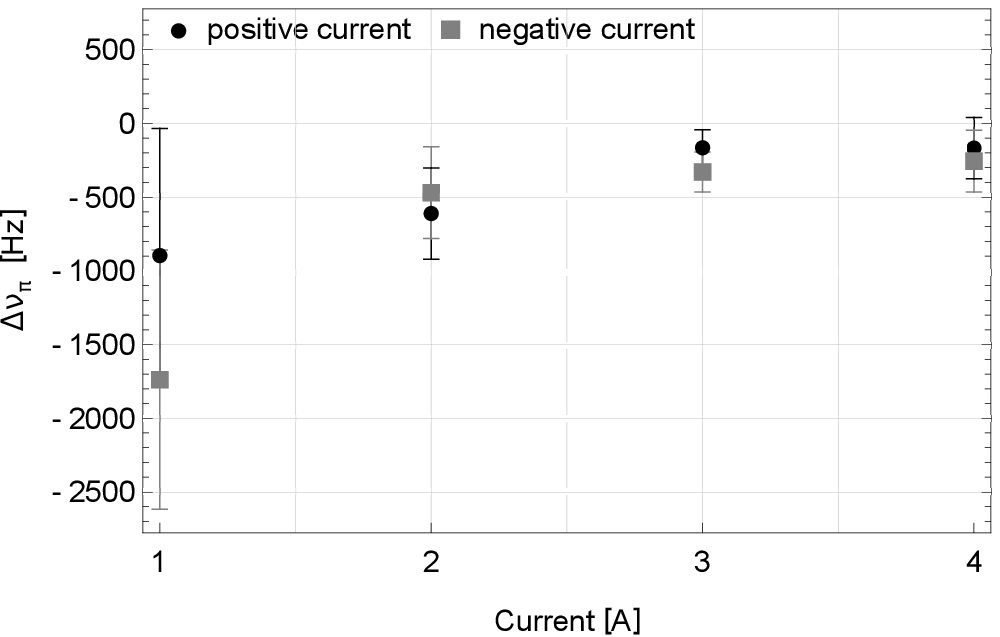}
\end{center}
\caption{Left: cavity for 1.42 GHz with magnet coils to produce \Bext\  and thre layers of magnetic shielding. Right: preliminary results of $\Delta$\nup\ for opposite directions of \Bext\ and various currents during a test run in 2021. }
\label{wid:fig1}
\end{figure}

A polarized H beam with a velocity corresponding to a temperature of 50 K has been built and used for a first measurement of the GS-HFS by extrapolating \nus\ to zero external magnetic field \Bext, yielding an accuracy of 2.7 ppb \cite{Diermaier:2017} and allowing to estimate that 8000 \Hbar\ atoms of similar velocity will be needed to perform a measurement with 1 ppm precision. Using an improved coil system to achieve higher homogeneity (cf. Fig.~\ref{wid:fig1} left) the $\pi$-transition could also be observed \cite{Widmann:2019} and in 2021 a measurement campaign was started to compare \nup\ for opposite horizontal directions of \Bext. 

The difference $\Delta$\nup\ = \nup(\Bext) $-$ \nup($-$\Bext) is sensitive to a set of non-minimal SME coefficients \cite{Kostelecky:2015}. Fig.~\ref{wid:fig1} right shows the result of a test beam in 2021 for four values of the coil current, where $I=1$ A $\widehat{=}$ \Bext$=2.3$ Gauss. Due to a problem with our frequency reference at that time the value for different currents is not the same, but the values at the same current for $+I$ and $-I $ overlap within the errors bars. At a typical error of 200 Hz (1$\sigma$) for $\Delta$\nup, a sensitivity of $6\times10^{-20}$ GeV for the coefficients  $\cnr{w}{kjm}$, $\anr{w}{kjm}$, $\gqPnr{w}{kjm}$, $\HqPnr{w}{kjm}$ ($w$: e, p, $k=2q$: power of the fermion momentum $|\boldmath{p}|$, $ij$: indices of spherical harmonics)\cite{Kostelecky:2015} can be expected from a high-statistics run performed in 2022. The error is dominated by the usage of \nus\ as a proxy for the magnetic field, which has a very weak dependence \nus(\Bext) at the low values of \Bext\ used in the experiment. An improvement of 1--2 orders of magnitude can be expected from using magnetometry to assure the same magnitude for opposite directions of \Bext.  

\section{HFS spectroscopy of deuterium}

Deuterium containing the lightest nucleus shows additional features to H and \Hbar. Since in the non-minimal SME the Hamiltonean is expanded in powers of the fermion momentum $\boldmath{p}$, the shifts of HFS levels depend on coefficients for the proton and neutron multiplied by the expectation value $\langle \boldmath{p}_{pd}^{2q} \rangle$ of the relative momentum $\boldmath{p}_{pd}$ of p vs. n in the d nucleus to the power $2q$. This leads to a $10^9$-times sensitivity enhancement for $k=2$ and $10^{18}$-times for $k=4$ \cite{Kostelecky:2015}. 

\begin{figure}
\begin{center}
\includegraphics[width=1.9in]{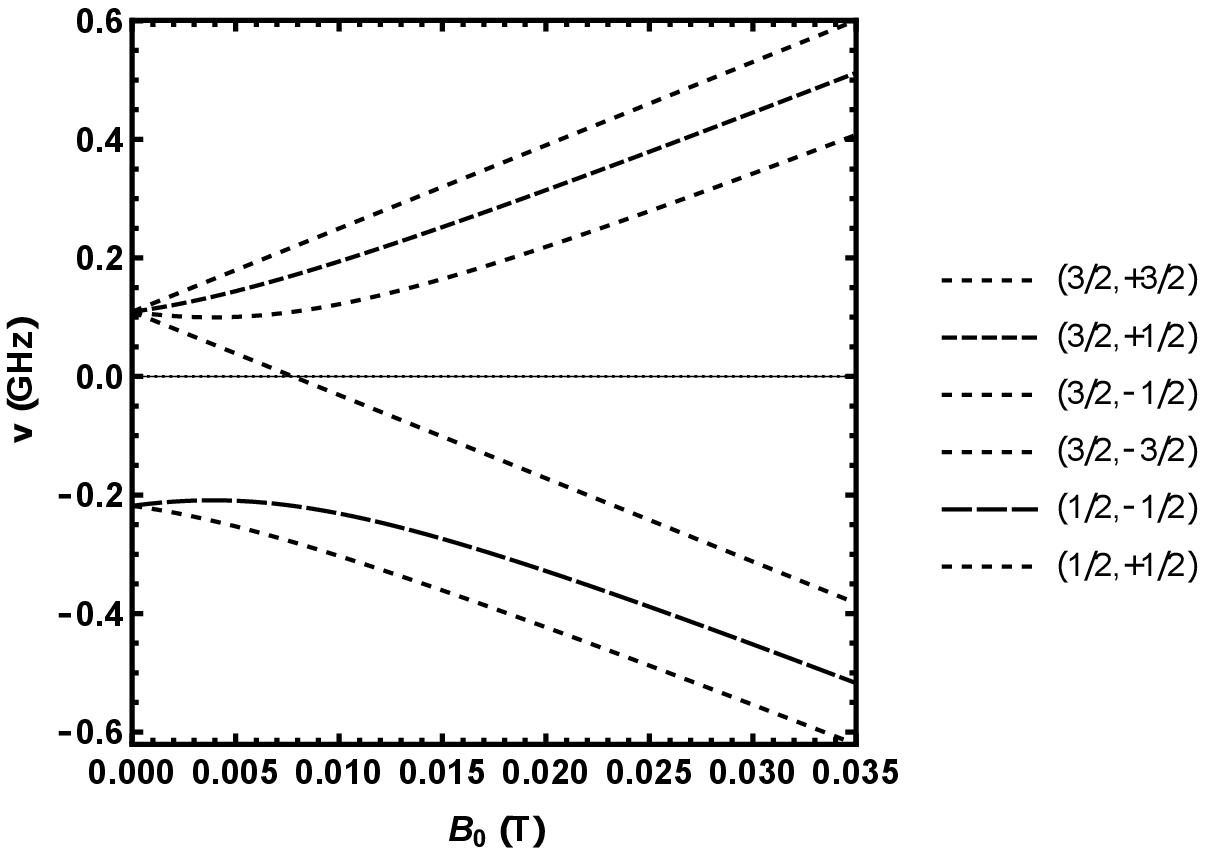} \hspace*{3mm}
\includegraphics[width=2.2in]{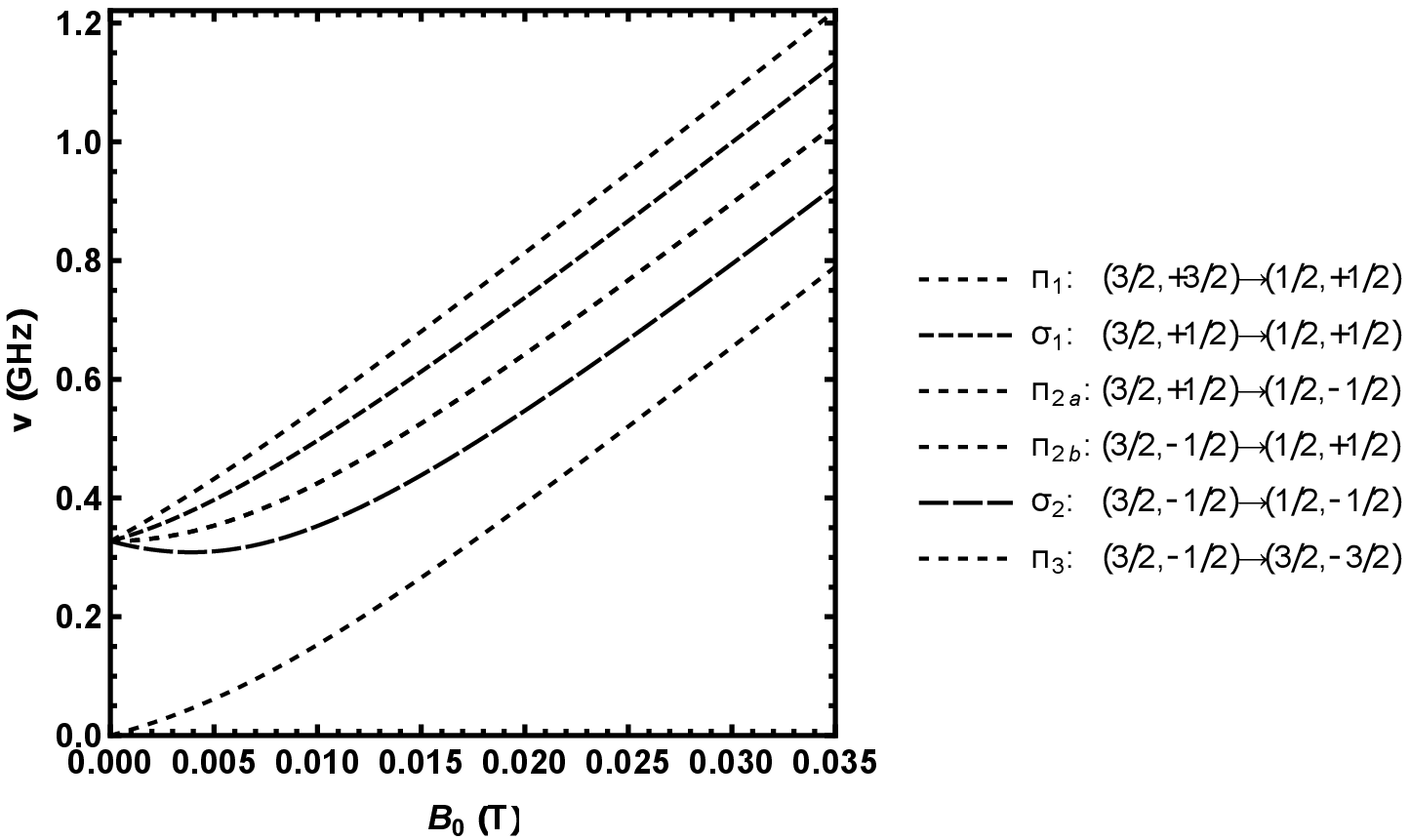}
\end{center}
\caption{Breit-Rabi diagram for deuterium (left) and transitions within the hyperfine levels (right). Values in brackets stand for the quantum numbers $(F,M_F)$.}
\label{wid:fig2}
\end{figure}

Fig.~\ref{wid:fig2} shows the Breit-Rabi diagram (left) and hyperfine transitions (right) for deuterium. In contrast to H, here the $\sigma_{1,2}$ transitions are sensitive to SME effects \cite{Kostelecky:2015}. The $\sigma_1$-transition $(3/2,+1/2)\rightarrow(1/2,+1/2)$ was used for the D maser \cite{Wineland:72}. We propose to use the $\sigma_2$-transition $(3/2,-1/2)\rightarrow(1/2,-1/2)$ which has a minimum of $\sim 309$ MHz at \Bext = 38.9 Gauss and to search for siderial variations at that point.

\begin{figure}
\begin{center}
\includegraphics[width=2.5in]{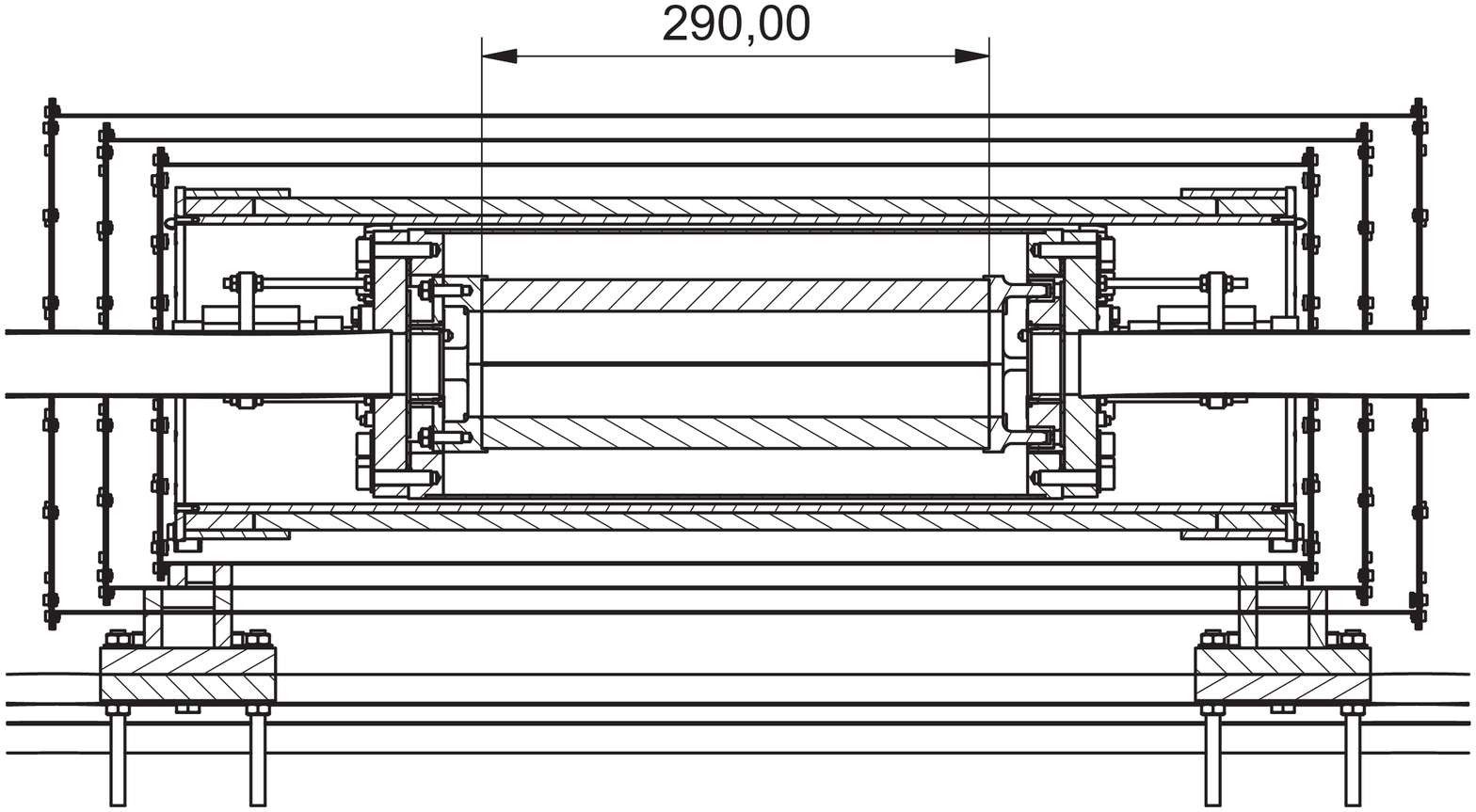} \hspace*{3mm}
\includegraphics[width=1.8in]{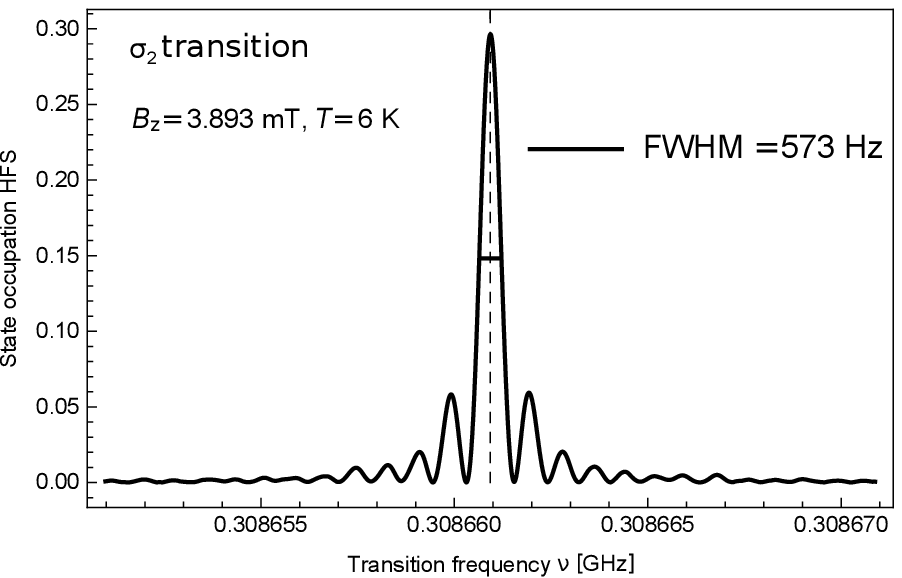}
\end{center}
\caption{Breit-Rabi diagram for deuterium (left) and transitions within the hyperfine levels (right). Values in brackets stand for the quantum numbers $(F,M_F)$. The length of the split ring resonator of 290 mm is indicated in the left figure.}
\label{wid:fig3}
\end{figure}

An experiment is being prepared using a deuterium beam at a temperature of $\sim 6$ K as was recently demonstrated for H \cite{Cooper:20}. For generating the 309 MHz radiation a split-ring resonator \cite{Hardy:81} was chosen for its compactness and independence of the resonance frequency from its length. The $\sigma$-transition allows to use a solenoid magnet with \Bext\ pointing in beam direction. Using the setup under construction shown in Fig.~\ref{wid:fig3} (left), a resonance line of less than 600 Hz width is expected as shown by the simulation in Fig.~\ref{wid:fig3} (right). With high statistics available in matter beams, a sub-Hz resolution should be achievable.

\section{Conclusions}

In-beam hyperfine spectroscopy offers sensitive tests of CPT and Lorentz invariance for H, \Hbar, and D. The ASACUSA  collaboration has made significant improvements to their mixing procedures which will allow spectroscopy to commence soon. For hydrogen, analysis is ongoing on data taken on the orientation dependence of an externally applied magnetic field and results should become available within less than one year. An experiment to measure siderial variations of HFS transitions in deuterium is being prepared and is expected to be performed within the next year.

\section*{Acknowledgments}

The author is indebted to the members of  ASACUSA for the continuous collaboration. Fruitful discussions with V.A. Kosteleck{\`y}, A.J. Vargas, and R. Lehnert are acknowledged. This work was supported by the European Research Council under European Union’s Seventh
Framework Programme (FP7/2007-2013) / ERC Grant Agreement (291242); the Austrian
Ministry of Science and Research, Austrian Science Fund (FWF) W1252-N27, P 32468 and P 34438;
the JSPS KAKENHI Fostering Joint International Research B 19KK0075; the Grant-in-Aid
for Scientific Research B 20H01930; Special Research Projects for Basic Science of RIKEN;
Università di Brescia and Istituto Nazionale di Fisica Nucleare; and the European Union’s
Horizon 2020 research and innovation programme under the Marie Skłodowska-Curie grant
agreement No 721559.

%\bibliographystyle{EBW-etal-it}
%\bibliography{hbar-physrep,D-rabi}

\begin{thebibliography}{10}

\bibitem{Widmann:22}
E.~Widmann,  
Phys. Part. Nucl.  \textbf{53} (2022) 790, 
arXiv:2111.04056 [hep-ex]
%Hyperfine spectroscopy of antihydrogen, hydrogen, and deuterium
  (2021).

\bibitem{Phillips:01}
D.~F. Phillips, M.~A. Humphrey, E.~M. Mattison, R.~E. Stoner, R.~F.~C. Vessot
  and R.~L. Walsworth, Phys. Rev. D \textbf{63} (2001) 111101.

\bibitem{Humphrey:2003}
M.~A. Humphrey, D.~F. Phillips, E.~M. Mattison, R.~F.~C. Vessot, R.~E. Stoner
  and R.~L. Walsworth, Phys. Rev. A \textbf{68} (2003) 063807.

\bibitem{Mohri:2003wu}
A.~Mohri and Y.~Yamazaki, Europhys. Lett. \textbf{63} (2003) 207.

\bibitem[]{KolbingerEtAl2015}
B. Kolbinger \etal, Eur. Phys. J. \textbf{D} 75, 91. 


\bibitem{Hunter:22}
E.~D. {Hunter} \etal, in:
  EPJ Web of Conferences  \textbf{262}  (2022) 01007.

\bibitem{Radics:14}
B.~Radics, D.~J. Murtagh, Y.~Yamazaki and F.~Robicheaux, Phys. Rev. A \textbf{90} (2014)
  032704.

\bibitem{Diermaier:2017}
M.~Diermaier, C.~B. Jepsen, B.~Kolbinger, C.~Malbrunot, O.~Massiczek,
  C.~Sauerzopf, M.~C. Simon, J.~Zmeskal and E.~Widmann, Nat. Commun. \textbf{8}
  (2017) 15749.

\bibitem{Widmann:2019}
E.~Widmann \etal, Hyperfine Interact. \textbf{240} (2019) 5.

\bibitem{Kostelecky:2015}
V.~A. Kosteleck\'y and A.~J. Vargas, Phys. Rev. D \textbf{92} (2015) 056002.

\bibitem{Wineland:72}
D.~J. Wineland and N.~F. Ramsey, Phys. Rev. A \textbf{5} (1972) 821.

\bibitem{Cooper:20}
S.~F. Cooper, A.~D. Brandt, C.~Rasor, Z.~Burkley and D.~C. Yost, Rev. Sci. Instrum \textbf{91} (2020)(1) 013201.

\bibitem{Hardy:81}
W.~N. Hardy and L.~A. Whitehead, Rev. Sci. Instrum \textbf{52} (1981)(2)
  213.

\end{thebibliography}

\end{document}